\documentclass[runningheads]{llncs}
\usepackage[a4paper, margin=1.68in]{geometry}

\usepackage{amsmath}
\usepackage{amssymb}
\usepackage{algorithm}
\usepackage{algorithmic}
\usepackage{natbib}
\usepackage{graphicx}
\usepackage{url}
\usepackage{color}
\usepackage{comment}
\usepackage{todonotes}
\usepackage{hyperref}
\usepackage{longtable}
\usepackage{wrapfig}
\usepackage{multirow}

\usepackage{subcaption} % For subfigures
\usepackage{float} % For controlling figure placement
\usepackage[T1]{fontenc}

\usepackage{hyperref}

\begin{document}

\title{HEAL: Hierarchical Embedding Alignment Loss for Improved Retrieval and Representation Learning}% for RAG and LLM Systems}

% If the paper title is too long for the running head, you can set an abbreviated paper title here
\titlerunning{HEAL}

%\author{Anonymous Author}
%\author{Manish Bhattarai, Ryan Barron, Maksim Eren, Minh Vu , Vesselin Grantcharov, Ismael Boureima, Valentin Stanev, Cynthia Matuszek, Vladimir Valtchinov, Kim Rasmussen, Boian Alexandrov}

\author{Manish Bhattarai\inst{1}\orcidID{0000-0002-1421-3643} \and
Ryan Barron\inst{1} \and
Maksim Eren\inst{2} \and
Minh Vu\inst{1} \and
Vesselin Grantcharov\inst{1} \and
Ismael Boureima\inst{1} \and
Valentin Stanev\inst{3} \and
Cynthia Matuszek\inst{4} \and
 Vladimir Valtchinov\inst{5} \and
 Kim Rasmussen\inst{1} \and
 Boian Alexandrov\inst{1}
}

\institute{Theoretical Division, Los Alamos National Laboratory, Los Alamos, NM, USA \and
Analytics Divison, Los Alamos National Laboratory, Los Alamos, NM, USA \and
Department of Material Science \& Engineering, University of Maryland, College Park, MD, USA \and
Department of Computer Science, University of Maryland, Baltimore County, MD, USA \and
Department of Radiology, Brigham and Women's Hospital, Harvard medical School, Boston MA}

\authorrunning{Bhattarai et al.}

\maketitle

\begin{abstract}

Retrieval-Augmented Generation (RAG) enhances Large Language Models (LLMs) by integrating external document retrieval to provide domain-specific or up-to-date knowledge. The effectiveness of RAG depends on the relevance of retrieved documents, which is influenced by the semantic alignment of embeddings with the domain's specialized content. Although full fine-tuning can align language models to specific domains, it is computationally intensive and demands substantial data. This paper introduces \textbf{H}ierarchical \textbf{E}mbedding \textbf{A}lignment \textbf{L}oss (HEAL), a novel method that leverages hierarchical fuzzy clustering with matrix factorization within contrastive learning to efficiently align LLM embeddings with domain-specific content. HEAL computes level/depth-wise contrastive losses and incorporates hierarchical penalties to align embeddings with the underlying relationships in label hierarchies. This approach enhances retrieval relevance and document classification, effectively reducing hallucinations in LLM outputs. In our experiments, we benchmark and evaluate HEAL across diverse domains, including Healthcare, Material Science, Cyber-security, and Applied Maths. %The results demonstrate that HEAL significantly improves retrieval relevance and downstream classification performance, thereby enhancing the reliability and accuracy of LLM-based systems%\todo[]{If we find that it is better than full fine-tuning, mention here too}.

%This paper introduces the \textbf{Hierarchical Embedding Alignment Loss (HEAL)}, a novel method designed to incorporate hierarchical label structures into contrastive learning to efficiently align LLMs into specific domains, allowing improved retrival relevance and document classification which in return reduces hallucinations. HEAL computes level-wise contrastive losses and aggregates them with hierarchical penalties, aligning embeddings with hierarchical relationships. The method is evaluated on benchmark datasets and demonstrates significant improvements in capturing hierarchical semantics compared to existing approaches. We apply HEAL in the context of Retrieval-Augmented Generation (RAG) systems, fine-tuning embedding models to improve retrieval accuracy and reduce hallucinations in large language models (LLMs). Experimental results show that HEAL enhances retrieval relevance and downstream classification tasks, contributing to more reliable and accurate LLM outputs.

\end{abstract}

\section{Introduction}
%\footnote{Corresponding Author: \email{ceodspspectrum@lanl.gov}}

Large Language Models (LLMs), such as GPT-4 \citep{openai2023gpt4}, have demonstrated exceptional capabilities in natural language understanding and generation. However, LLMs are prone to \emph{hallucinations}, generating plausible but incorrect or nonsensical content \citep{ji2023survey}. Retrieval-Augmented Generation (RAG) frameworks \citep{lewis2020retrieval} mitigate this issue by integrating external knowledge through document retrieval, enhancing the factual accuracy of LLM outputs. A critical component of RAG systems is the embedding model used for document retrieval. Standard embedding models, however, often fail to capture the hierarchical and semantic relationships within domain-specific corpora, leading to suboptimal retrieval and, consequently, increased hallucinations. This issue is particularly pronounced in %high-stakes
domains with increased specificity such as Healthcare, Legal sytem, and Scientific research.%, where precise information retrieval is essential.

Corpus of documents for a specialized domain inherently exhibit a high degree of semantic coherence, presenting an opportunity to align embedding models for retrieving the most contextually relevant information. Hierarchical Non-negative Matrix Factorization (HNMF) \citep{eren2023semi} is a powerful technique for semantically categorizing documents into clusters that exhibit thematic coherence. By grouping documents into hierarchical clusters of supertopics and subtopics, HNMF provides a rich semantic categorization of the corpus, enabling a deeper understanding of document relationships. Leveraging this semantic knowledge in the form of hierarchical cluster labels, we can align embedding models to preserve hierarchical information within the embedding space. This alignment enhances the embeddings to capture both coarse-grained and fine-grained document similarities, improving contextual relevance in retrieval tasks and enabling better downstream capabilities.

To tackle the challenges of hallucination and suboptimal retrieval in RAG systems, we introduce the \textbf{Hierarchical Embedding Alignment Loss (HEAL)}, a refined extension of the Hierarchical Multi-label Contrastive Loss~\citep{zhang2022use}. HEAL leverages an improved hierarchical weighting scheme to align embeddings more effectively with the underlying hierarchical structure. By incorporating hierarchical label structures, HEAL fine-tunes embedding models to align with document clusters derived from HNMF. The method computes contrastive losses at each hierarchical level, combining them with depth-specific penalties to emphasize distinctions at higher levels of the hierarchy. 

Our contributions are summarized as follows:
\begin{enumerate}
    \item Introduce a refined contrastive learning framework, named HEAL, that incorporates hierarchical label structures to align embeddings with hierarchical document relationships.
    \item Integrate HEAL into RAG systems, fine-tuning embedding models to improve retrieval accuracy and reduce hallucinations in LLM outputs.
    \item Validate and benchmark HEAL through extensive experiments on domain-specific datasets from specialized scientific sub-domains of Healthcare, Material Science, Tensor Decomposition, and Cyber-security.
    \item Showcase significant improvements in retrieval relevance and downstream tasks compared to baseline method.
\end{enumerate}

%The rest of the paper is organized as follows. Section~\ref{sec:related} reviews related work, including contrastive learning, hierarchical representation learning, and RAG frameworks. Section~\ref{sec:method} provides a detailed explanation of the HEAL methodology, including rigorous mathematical formulations. Section~\ref{sec:experiments} presents experimental results and analysis, while Section~\ref{sec:conclusion} concludes the paper with a summary of contributions and future directions.

\section{Related Work} \label{sec:related}

%\subsection{Contrastive Learning}

Contrastive learning has become a cornerstone of representation learning, particularly in computer vision and natural language processing. Methods like SimCLR \citep{chen2020simple} and MoCo \citep{he2020momentum} have achieved state-of-the-art performance in unsupervised settings by learning representations that are invariant to data augmentations. In supervised contrastive learning, \citet{khosla2020supervised} extended the contrastive loss to utilize label information, improving performance on classification tasks. Similarly, the SciNCL framework employs neighborhood contrastive learning to capture continuous similarity among scientific documents, leveraging citation graph embeddings to sample both positive and negative examples~\cite{Ostendorff2022scincl}. However, these methods generally assume flat label structures and do not exploit hierarchical relationships.

%\subsection{Hierarchical Representation Learning}

Hierarchical classification has been studied extensively, with approaches such as hierarchical softmax \citep{goodman2001classes} and hierarchical cross-entropy loss \citep{deng2014large}. These methods aim to leverage hierarchical label structures to improve classification efficiency and accuracy. In the context of representation learning, \citet{deng2011hierarchical} introduced hierarchical semantic embedding, aligning image embeddings with WordNet hierarchies. More recent works, such as \citet{bertinetto2020making}, have explored hierarchical prototypes to capture hierarchical relationships.
\citet{zhang2022use} propose a hierarchical multi-label contrastive learning framework that preserves hierarchical label relationships through hierarchy-preserving losses. Their method excels in scenarios with hierarchical multi-label annotations, such as biological or product classifications. In contrast, our approach focuses on enhancing information retrieval to mitigate hallucinations.

%\subsection{Retrieval-Augmented Generation}

RAG frameworks combine retrieval models with generative models to enhance the factual accuracy of language generation \citep{lewis2020retrieval}. These systems rely heavily on the quality of the embeddings used for retrieval. Prior work has focused on improving retrieval through better indexing and retrieval algorithms \citep{karpukhin2020dense}, but less attention has been given to aligning embeddings with hierarchical document structures.

%\todo[inline]{we should add more relevant work here and compare to our method to show what we are improving}

\section{Method} \label{sec:method}

\begin{figure}[!ht]
    \centering
    \includegraphics[width=1\linewidth]{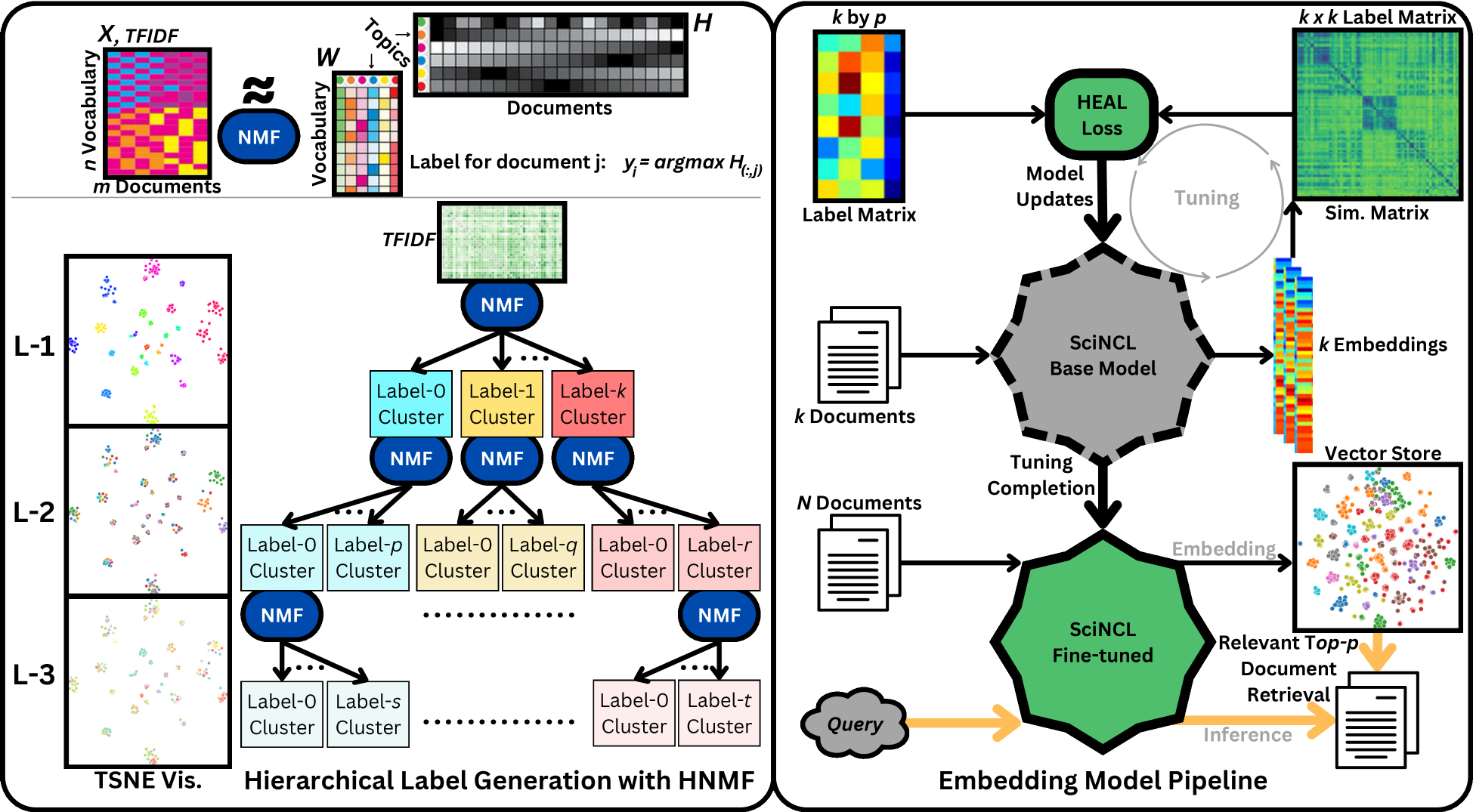}
    \caption{Overview of the HEAL-Based Embedding Model Alignment and Retrieval. \textit{The left side} illustrates hierarchical label generation using HNMF, where documents corresponding to a cluster from each preceding depth are converted into TFIDF matrices and further decomposed to extract sub-clusters. The TSNE visualizations highlighting cluster memberships in document embeddings. \textit{The right side} depicts fine-tuning of the SciNCL model using HEAL loss on generated embeddings and HNMF derived labels. Once trained, the aligned model computes a vector store from the corpus, enabling retrieval of the nearest $p$ documents for a given query embedding. }
    \label{fig:heal-overview}
    \vspace{-2mm}
\end{figure}

In this section, we propose an embedding alignment framework comprising hierarchical label extraction with HNMF, embedding alignment using HEAL, and retrieval with aligned embeddings as outlined in Figure~\ref{fig:heal-overview}.
%In this section, we introduce the \textbf{HEAL}, a loss function designed to incorporate hierarchical label information into contrastive learning. HEAL extends the Supervised Contrastive Loss \citep{khosla2020supervised} by computing contrastive losses at each hierarchical level and aggregating them with level-specific penalties. %This method aligns the embedding space with the hierarchical structure of the data, emphasizing discrepancies at higher levels to better reflect hierarchical label relationships.

\subsection{Hierarchical Document Clustering with HNMFk. } \label{hnmfk_para}
%\todo[]{we should add a brief description at caption of what we are showing in the figure \ref{fig:heal-overview} to help the reader understand what they are looking at}

Hierarchical Non-negative Matrix Factorization with automatic latent feature estimation (HNMFk)~\citet{eren2023semi} is an advanced technique for uncovering hierarchical patterns within document collections. It builds on traditional Non-negative Matrix Factorization (NMF)~\citet{vangara2021finding} by 
dynamically and automatically determining the optimal number of latent features at each level.
%also called automatic model determination. 
%This automatic determination of model parameters ensures that the number of document clusters is appropriately selected to avoid over-fitting (selecting too many clusters and introducing noise) or under-fitting (selecting too few clusters and blending distinct documents or topics).
Effective contrastive learning relies on well-separated document cluster labels to align embeddings effectively. HNMFk's ability to automatically balance stability and accuracy using a bootstrap approach enhances the quality of clustering results. %This capability improves both the interpretability of the hierarchical structure and the precision of the clustering process. 
In this work, we utilize the publicly available HNMFk implementation from the TELF library \footnote{TELF is available at \url{https://github.com/lanl/T-ELF}}.

Given a Term Frequency-Inverse Document Frequency (TF-IDF) matrix \(\mathbf{X} \in \mathbb{R}^{n \times m}\), where \(n\) represents the vocabulary size and \(m\) denotes the number of documents, HNMFk performs a sequence of matrix factorizations across hierarchical levels to capture the nested structure of topics. At each level \(l\), the factorization is expressed as \(\mathbf{X} \approx \mathbf{W}^{(l)} \mathbf{H}^{(l)}\), where \(\mathbf{W}^{(l)} \in \mathbb{R}^{n \times k_l}\) is the basis matrix representing latent topics, and \(\mathbf{H}^{(l)} \in \mathbb{R}^{k_l \times m}\) is the coefficient matrix quantifying the contribution of each topic to the composition of documents. Here, \(k_l\) is the number of topics at level \(l\), which is determined automatically through stability analysis~\cite{vangara2021finding}. This analysis involves bootstrapping the data to create resampled versions of the TF-IDF matrix, applying NMF across a range of \(k\) values, and evaluating the stability of clusters across the resampled datasets. The optimal \(k_l\) is selected as the value that produces the most consistent clustering results, indicating a robust underlying structure in the data.

To construct hierarchical labels for each document, the coefficient matrix \(\mathbf{H}^{(l)}\) is used to determine topic assignments. For each level \(l\), the topic for document \(i\) is identified by selecting the index of the maximum value in the corresponding column of \(\mathbf{H}^{(l)}\), expressed as \(y_i^{(l)} = \arg\max_k \mathbf{H}^{(l)}_{k, i}\). The hierarchical label for document \(i\) is then formed by aggregating the topic assignments across all levels, resulting in \(\mathbf{y}_i = (y_i^{(0)}, y_i^{(2)}, \ldots, y_i^{(L-1)})\). Here, \(L\) is the total number of hierarchical levels, or hierarchical depth that is the number of NMFk operations from the first one to the leaf. \(y_{i}^l\) is the label of sample \(i\) at level \(l\), with \(l = 0\) corresponding to the \emph{shallowest}(most general or root node)  level and \(l = L-1\) to the \emph{deepest} (most fine-grained, or leaf node) level.

\subsection{Hierarchical Multilevel Contrastive Loss (HEAL)} \label{heal}
%citing HNMFk 

Upon the unsupervised data decomposition with HNMFk, the datasets have clusters with hierarchical structures. To incorporate such structures, we propose the HEAL, which extends supervised contrastive loss~\citep{khosla2020supervised} by introducing level-wise contrastive losses and aggregating them with level-specific penalties.

\subsubsection{Level-wise Contrastive Loss}
 For a batch of \(N\) samples \(\{(\mathbf{x}_i, \mathbf{y}_i)\}_{i=1}^N\), where \(\mathbf{x}_i \in \mathbb{R}^d\) is the input and $\mathbf{y}_i\in \mathbb{R}^L$ is the hierarchical cluster label, we obtain normalized embeddings \(\{\mathbf{h}_i\}_{i=1}^N\) using an encoder network \(f_{\theta}(\cdot)\):

\begin{equation}
    \mathbf{h}_i = \frac{f_{\theta}(\mathbf{x}_i)}{\|f_{\theta}(\mathbf{x}_i)\|_2}, \quad \mathbf{h}_i \in \mathbb{R}^d.
\end{equation}

%Based on~\ref{hnmfk_para}, each sample \(i\) is associated with a hierarchical label vector \(\mathbf{y}_i = (y_i^{(0)}, y_i^{(2)}, \ldots, y_i^{(L-1)})\).
For a given level \(l\), the set of positive samples for sample \(i\) is:

\begin{equation}
\label{eq:pos_set}
    P(i, l) = \{ p \mid \mathbf{y}_{p}^{l} = \mathbf{y}_{i}^{l}, p \neq i \}.
\end{equation}

The contrastive loss at level \(l\) for sample \(i\) is:

\begin{equation} \label{eq:level_loss}
    \mathcal{L}_{i,l} = \frac{-1}{|P(i, l)|} \sum_{p \in P(i, l)} \log \frac{\exp\left( \mathbf{h}_i^\top \mathbf{h}_p / \tau \right)}{\sum_{a=1}^N \exp\left( \mathbf{h}_i^\top \mathbf{h}_a / \tau \right)}.
\end{equation}

If \(P(i, l)\) is empty (i.e., no positive samples at level \(l\) for \(i\)), \(\mathcal{L}_{i,l}\) is excluded from the total loss.

\subsubsection{Aggregating Level-wise Losses with Penalties}

To prioritize discrepancies at shallower levels, we assign penalties \(\lambda_l\) to each level \(l\), where shallower levels have higher penalties. The penalties are defined as:

\begin{equation} \label{eq:penalty}
    \lambda_l = \frac{2^{L-l-1}}{\sum_{k=0}^{L-1} 2^{k}} = \frac{2^{L-l-1}}{2^L - 1}.
\end{equation}
The penalties \(\lambda_l\) satisfy:

\begin{enumerate}
    \item \(\lambda_{l} > \lambda_{l+1}\) for \(l = 0,1,...,L-2 \), i.e., penalties decrease for deeper levels.
    \item \(\sum_{l=0}^{L-1} \lambda_l = 1\), i.e., the penalties are normalized.
\end{enumerate}
The total HEAL loss is then:

\begin{equation} \label{eq:HEAL_loss}
    \mathcal{L}_{\text{HEAL}} = \frac{1}{N} \sum_{l=0}^{L-1} \lambda_l \sum_{i=1}^N \mathcal{L}_{i,l}.
\end{equation}

\begin{algorithm}[H]
\caption{Computation of HEAL Loss} \label{alg:HEAL}
\begin{algorithmic}[1]
\REQUIRE Mini-batch \(\{(\mathbf{x}_i, \mathbf{y}_i)\}_{i=1}^N\), temperature \(\tau\), number of levels \(L\)
\STATE Compute embeddings: \(\mathbf{h}_i = f_{\theta}(\mathbf{x}_i) / \|f_{\theta}(\mathbf{x}_i)\|_2\)
\STATE Initialize total loss: \(\mathcal{L}_{\text{HEAL}} \leftarrow 0\)
\FOR{\(l = 0\) to \(L-1\)}
    \STATE Compute penalty \(\lambda_l\) using Eq.~\eqref{eq:penalty}
    \FOR{\(i = 1\) to \(N\)}
        \STATE Determine positive set \(P(i, l)\) using Eq.~\eqref{eq:pos_set}
        \IF{\(|P(i, l)| > 0\)}
            \STATE Compute \(\mathcal{L}_{i,l}\) using Eq.~\eqref{eq:level_loss}
            \STATE Update total loss: \(\mathcal{L}_{\text{HEAL}} \leftarrow \mathcal{L}_{\text{HEAL}} + \lambda_l \mathcal{L}_{i,l}\)
        \ENDIF
    \ENDFOR
\ENDFOR
\RETURN \(\mathcal{L}_{\text{HEAL}}\)
\end{algorithmic}
\end{algorithm}
%\subsubsection{Properties of Penalties}

%\subsection{Implementation Details}

%\subsubsection{Numerical Stability and Normalization}
\begin{comment}

To prevent scale issues, embeddings are normalized to unit vectors:

\begin{equation}
    \mathbf{h}_i = \frac{f(\mathbf{x}_i)}{\|f(\mathbf{x}_i)\|_2}.
\end{equation}

For numerical stability, logits are adjusted by subtracting the maximum logit:

\begin{equation}
    z_{i,a} = \frac{\mathbf{h}_i^\top \mathbf{h}_a}{\tau} - \max_{k} \left( \frac{\mathbf{h}_i^\top \mathbf{h}_k}{\tau} \right).
\end{equation}

%\subsubsection{Masking Self-Contrast}

We exclude self-contrast terms by ensuring \(i \neq p\) in \(P(i, l)\).
To account for varying level distributions in mini-batches, penalties are dynamically normalized:

\begin{equation} \label{eq:dynamic_pen}
    \lambda_l^{\text{batch}} = \frac{2^{L-l-1}}{\sum_{k \in \mathcal{L}_{\text{batch}}} 2^{L-k-1}}
,
\end{equation}

where \(\mathcal{L}_{\text{batch}}\) is the set of levels with non-empty positive sets.
\end{comment}
%\subsection{Optimization Algorithm}

Algorithm~\ref{alg:HEAL} outlines the computation of \(\mathcal{L}_{\text{HEAL}}\) for a mini-batch.

\subsection{Fine-tuning Embedding Models with HEAL for RAG}

To enhance retrieval performance in RAG systems, we fine-tune the embedding model to align with the hierarchical structure of the document corpus. Given a specialized document corpus, we first apply HNMFk (as described in Section~\ref{hnmfk_para}) to the corresponding TF-IDF matrix \(\mathbf{X}\) producing hierarchical cluster labels \(\mathbf{y}_i = (y_i^{(0)}, y_i^{(2)}, \ldots, y_i^{(L-1)})\) for each document \(i\). Next, we generate embeddings from each document $x_i$ using a pretrained embedding model $f_{\theta}(.)$. The embedding model is initialized with pre-trained weights and produces normalized embeddings \(\mathbf{h}_i \in \mathbb{R}^d\) for document \(i\).
\begin{comment}

\[
\mathbf{h}_i = \frac{f_{\theta}(\mathbf{x}_i)}{\|f_{\theta}(\mathbf{x}_i)\|_2}.
\]
\end{comment}
To align embeddings with the hierarchical structure, we optimize the HEAL presented in \ref{heal}.

The embedding model is trained by minimizing \(\mathcal{L}_{\text{HEAL}}\) using gradient-based optimization:
\[
\mathbf{\theta}^* = \arg \min_{\mathbf{\theta}} \mathcal{L}_{\text{HEAL}},
\]
where \(\mathbf{\theta}\) are the parameters of the embedding model \(f_\theta(\cdot)\).

After fine-tuning, the updated embeddings \(\mathbf{h}_i = f_{\theta^*}(\mathbf{x}_i)\) are used to replace the initial embeddings in the vector store. During inference, a query \(\mathbf{q}\) is embedded using \(f_{\theta^*}(\cdot)\) as \(\mathbf{h}_q = f_{\theta^*}(\mathbf{q})\), and  
retrieves top $p$ documents based on cosine similarity:
\[
\text{Similarity}(\mathbf{q}, \mathbf{x}_i) = \frac{\mathbf{h}_q^\top \mathbf{h}_i}{\|\mathbf{h}_q\| \|\mathbf{h}_i\|}.
\]

To maximize retrieval performance in RAG systems, it is essential to align the query embeddings with the hierarchically aligned document embeddings. Since queries are typically shorter and may not capture the full semantic richness of the documents, we need to semantically align queries and documents in the embedding space. To achieve this, we generate question-answer (Q\&A) pairs  using a language model (e.g., LLaMA-3.1 70B) for each document and leverage HEAL to jointly align both query and document embeddings during training. 
For each document \(\mathbf{x}_i\), we generate a set of queries \(\{\mathbf{q}_{i,k}\}_{k=1}^{K_i}\), where \(K_i\) is the number of queries generated for document \(i\). Each query \(\mathbf{q}_{i,k}\) is associated with the same hierarchical labels \(\mathbf{y}_i\) as its source document \(\mathbf{x}_i\), since it is derived from the content of \(\mathbf{x}_i\).We extend the HEAL framework to include both documents and queries by defining a unified set of samples:

\[
\mathcal{S} = \{\mathbf{x}_1, \dots, \mathbf{x}_N\} \cup \{\mathbf{q}_{i,k} \mid i = 1, \dots, N; \, k = 1, \dots, K_i\}.
\]

Each sample \(\mathbf{s}_j \in \mathcal{S}\) has an associated hierarchical label \(\mathbf{y}_j\), where:

\[
\mathbf{y}_j = \begin{cases}
\mathbf{y}_i, & \text{if } \mathbf{s}_j = \mathbf{x}_i \text{ (document)}; \\
\mathbf{y}_i, & \text{if } \mathbf{s}_j = \mathbf{q}_{i,k} \text{ (query generated from document } \mathbf{x}_i).
\end{cases}
\]
Based on this dataset, the HEAL is leveraged to finetune the embedding model~\label{heal}.
\section{Experiments} \label{sec:experiments}

\subsection{Datasets}

We evaluate our method on datasets specifically constructed from scientific publications in the domains of Material Science, Medicine, Tensor Decomposition, and Cybersecurity. To construct our datasets, we leveraged the Bibliographic Utility Network Information Expansion (BUNIE) method, a machine learning-based approach that integrates subject-matter expertise in a human-in-the-loop framework \cite{10460022}.
For completeness, we briefly summarize the BUNIE approach in this paper. BUNIE begins with a small core corpus of documents selected by subject-matter experts (SMEs). From this starting point, it constructs a citation network to identify additional relevant documents, leveraging BERT based text embeddings to assess semantic similarity. Through iterative cycles of dataset expansion and pruning—guided by embedding visualization, topic modeling, and expert feedback—the method ensures the corpus is both comprehensive and domain-specific. %This dynamic refinement process enhances the dataset’s relevance while maintaining scalability, making it particularly suited for generating targeted scientific corpora. 
We apply this procedure to each scientific domain with guidance from SMEs, who provide target keywords/phrases and/or a core set of papers relevant to the sub-topic of interest within the domain. Using this knowledge base, we employ BUNIE to expand the dataset from the initial core papers to a larger collection of domain-specific documents. %\todo[inline]{should we do this at the appendix?} This process is repeated across the following domains to evaluate HEAL:

% \todo[inline]{Ryan, Could you also provide document counts for following dataset}
\begin{enumerate}
    \item \textbf{Material Science}: %A collection of 46,862 Material research articles about 73 Transition metal dichalcogenides (TMD) materials. TMDs combine transition-metal and chalcogen atoms (S, Se, or Te). These compounds have emerged as promising class of materials in several distinct areas of physics. Their layered structure similar to graphite makes them excellent solid lubricants, especially useful in extreme conditions where traditional oil-based lubricants cannot be applied. TMD also exhibit various quantum phases at low temperature, in particular superconductivity and charge density waves. Atomically thin layers of TMD materials can can synthesized with high precision and demonstrate remarkable tunability of their properties. These two-dimensional materials have a host of potential applications in fields such as spintronics, optoelectronics, energy harvesting, and flexible electronics. 
    A collection of 46,862 scientific articles, which explore 73 Transition Metal Dichalcogenides (TMD) compounds, combining transition-metal and chalcogen atoms (S, Se, or Te). With a layered structure similar to graphite, TMDs excel as solid lubricants and exhibit unique quantum phases like superconductivity and charge density waves. Their atomically thin layers offer tunable properties, with applications in spintronics, optoelectronics, energy harvesting, batteries, and flexible electronics.
    \item \textbf{Healthcare}: %A dataset of 9,639 peer-reviewed research papers about Pulmonary Hypertension (PH), a rare hemodynamic condition marked by elevated pulmonary arterial pressure, leading to right heart strain and reduced oxygen delivery (Galie et al., 2021). The World Health Organization (WHO) classifies PH into five groups: Group 1 (pulmonary arterial hypertension), Group 2 (PH due to left heart disease), Group 3 (PH due to lung diseases or hypoxia), Group 4 (chronic thromboembolic pulmonary hypertension), and Group 5 (PH with unclear mechanisms) (Gonzalez et al., 2021). In the U.S., PH prevalence is estimated at 15-25 cases per million, with pulmonary arterial hypertension (PAH) being a significant subset (Humbert et al., 2021). Treatment options include endothelin receptor antagonists, phosphodiesterase-5 inhibitors, soluble guanylate cyclase stimulators, and prostacyclin analogs, aimed at improving symptoms and prolonging survival (Tuder et al., 2021). Prognosis varies by etiology; untreated PAH has a median survival of less than three years.
    A collection of 9,639 scientific articles, which examine Pulmonary Hypertension (PH) disease - a rare condition causing elevated pulmonary arterial pressure, right heart strain, and reduced oxygen delivery. The WHO classifies PH into five groups based on causes, including pulmonary arterial hypertension (PAH), which has a prevalence of 15-25 cases per million in the U.S. Treatments such as endothelin receptor antagonists and prostacyclin analogs aim to improve symptoms, but prognosis varies, with untreated PAH having a median survival of less than three years.
 %PH References:
%Galie, N., et al. (2021). “Updated treatment algorithm for pulmonary arterial hypertension.” *Journal of the American College of Cardiology*, 77(16), 2140-2157.
%Gonzalez, A., et al. (2021). “Pulmonary hypertension: epidemiology, mechanisms, and management.” *Nature Reviews Cardiology*, 18(4), 247-267.
%Humbert, M., et al. (2021). “Pulmonary arterial hypertension: pathophysiology and clinical management.” *European Respiratory Journal*, 57(5), 2001761.
%Tuder, R. M., et al. (2021). “Pathobiology of pulmonary arterial hypertension.” *Journal of the American College of Cardiology*, 77(5), 678-693.
    \item \textbf{Applied Mathematics:} 
    %A collection of 4,624 research articles on tensor decomposition, a technique for breaking down tensors (multi-dimensional arrays) into simpler components. Common methods like Canonical Polyadic Decomposition, Tucker Decomposition, and tensor networks - like Tensor-Train (TT) decompositions demonstrated extraordinary efficiency for big-data representation, data compression, and data-analytics. Tensor decomposition is applied in many areas, such as, recommender systems, image processing, knowledge graph embedding, data-completion, signal processing, etc., where extra-large data challenges present themselves. The collection of research articles we investigated targets specifically solving Partial Differential Equations (PDEs) using TT networks. Tensor networks, as a highly efficient representation of high-dimensional data, enable a super-compression and efficient handling of complex PDEs. This emerging mathematical approach is particularly advantageous to mitigate the curse of dimensionality, resulting in reducing significantly the computational costs and memory usage, while maintaining a high accuracy of numerical solution. Tensor network-based PDE solvers hold significant promise for advancing scientific computing, enabling breakthroughs in fields such as material science, climate modeling, and engineering design optimization.
    % \todo[inline]{Ryan and Boian pls. provide details}
    A collection of 4,624 scientific articles, which explore tensor network techniques, such as Tensor-Train (TT) decomposition, which recently emerged as a powerful mathematical tool for solving large-scale Partial Differential Equations (PDEs). Tensor network PDE solvers efficiently manage high-dimensional data by mitigating the curse of dimensionality, drastically reducing computational costs and memory usage while maintaining high solution accuracy. These advancements hold significant promise for breakthroughs in scientific computing, including material science, climate modeling, and engineering design optimization.
    \item \textbf{Cyber-security}: We created a dataset of 8,790 scientific publications focusing on the application of tensor decomposition methods in cybersecurity and ML techniques for malware analysis. This dataset serves as a knowledge base covering topics for cyber-security such as ML-based anomaly detection, malware classification, novel malware detection, uncertainty quantification, real-world malware analysis challenges, tensor-based anomaly detection, malware characterization, and user behavior analysis.
\end{enumerate}

\subsection{Experimental Setup}

For training, we used the Adam optimizer with a learning rate of $10^{-5}$, a batch size of $128$, and early stopping based on validation performance with a patience of $5$ epochs. The experiments were conducted on a high-performance computing cluster, with each node equipped with $4$ NVIDIA GH200 GPUs. Document metadata, comprising the title and abstract combined, were used as input. Hierarchical labels were generated using HNMF with dataset-specific factorization depths: Material Science (depth 3), Healthcare (depth 4), Applied Mathematics (depth 3), and Cybersecurity (depth 3). HEAL loss was applied with a temperature parameter of $0.07$. The embedding base model, SciNCL~\cite{Ostendorff2022scincl}, was chosen for its robust contrastive pretraining on scientific documents, serving as a strong baseline for fine-tuning.
The data was split into $60\%$ training, $20\%$ validation, and $20\%$ test sets, with early stopping monitored on the validation set. Evaluation metrics were reported on the test set, while Q\&A retrieval analysis used the entire dataset (train + validation + test) for constructing the vector store.

The efficacy of the RAG system was evaluated at two levels. \textit{First}, we characterized the embeddings on document-level tasks, including hierarchical classification, retrieval, and hallucination measurement. For hierarchical classification, we used a hierarchical classifier applying random forests to each node~\citep{miranda2023hiclass}. The classifier is trained on embeddings corresponding to train dataset and evaluated against the test set. We perform this for embeddings derived from aligned and unaligned embedding model. Retrieval performance was assessed by measuring whether retrieved documents belonged to the same hierarchical class as the query document. Hallucination likelihood was evaluated based on the retrieval of incorrect documents for a given query. %This standard evaluation pipeline aligns with embedding model benchmarks in vision and NLP tasks.
\textit{Second}, we evaluated the performance of the embedding model within a RAG framework. To support retrieval and hallucination analysis, we used the LLaMA-3.1 70B model to generate $10$ Q\&A pairs per document using abstracts as input, providing a robust test for embedding alignment and retrieval capabilities. Next, we leveraged the questions as queries to the embedding model to retrieve the best metadata and assessed whether the model retrieved the exact document that generated the query during Q\&A analysis, as well as the rank of the returned document within the top 10 results. Furthermore, the retrieved documents were augmented with LLaMA-3.1 70B LLM to generate responses, with hallucinations evaluated based on response accuracy and relevance.

%\subsubsection{Baseline Methods}
Given the specialized nature of our dataset and the requirement for hierarchical labels, fine-tuning is essential. Comparing our method to approaches that do not leverage hierarchical labels is inequitable, as they are inherently less effective for this task. Our approach simplifies training by eliminating HEAL loss hyperparameter tuning, unlike HiMulCon~\cite{zhang2022use}, which requires extensive tuning of penalty parameters for optimal results. While HiMulCon focuses on root-level classification in vision datasets, our method aligns embeddings across all hierarchical depths. We optimize hierarchical metrics such as classification, retrieval, and hallucination indirectly through the HEAL loss, ensuring a robust alignment with the hierarchical structure.

For these reasons, we evaluate the performance of HEAL using the baseline model SciNCL, both without and with hierarchical alignment on our diverse specialized datasets.
\begin{comment}
\begin{itemize}
    \item \textbf{Standard Contrastive Learning (SCL)}: Fine-tuning using the standard Supervised Contrastive Loss \citep{khosla2020supervised}.
    \item \textbf{Hierarchical Cross-Entropy (HCE)}: Fine-tuning using hierarchical cross-entropy loss.
    \item \textbf{No Fine-tuning}: Using pre-trained embedding model without % additional 
    fine-tuning.
\end{itemize}
\end{comment}
We evaluate performance using  hierarchical metrics to capture nuances of hierarchical label structures in retrieval, classification, and hallucination assessments as presented in Table~\ref{tab:hierarchical_metrics} .

\begin{table}[!ht]
\centering
\begin{tabular}{|l|l|l|}
\hline
\textbf{Metric}                        & \textbf{Formula}                                                                                                  & \textbf{Description}                                                                 \\ \hline
\textbf{Hierarchical}        & $\text{Relevance}(q, r) = $                                  & Average label match \\ \textbf{Relevance} & $\frac{1}{L} \sum_{l=0}^{L-1} \delta(y_{q}^{l}, y_{r}^{l})$ & across hierarchy levels\\                                \hline
     & &   Fraction of                                                        \\  \textbf{Hierarchical}  & $\frac{1}{k} \sum_{i=1}^k \text{Relevance}(q, r_i)$  & hierarchically relevant \\ \textbf{Precision@k}&& documents among top $k$.                \\ \hline
       & & Fraction of \\ \textbf{Hierarchical} &$\frac{\sum_{i=1}^k \text{Relevance}(q, r_i)}{\sum_{r \in \text{Relevant}(q)} \text{Relevance}(q, r)}$& hierarchically relevant \\ \textbf{Recall@k}&& documents retrieved.                    \\ \hline
&& Discounted gain based\\ \textbf{Hierarchical}          & $\frac{\sum_{i=1}^k \frac{2^{\text{Relevance}(q, r_i)} - 1}{\log_2(i + 1)}}{\sum_{i=1}^k \frac{2^{\text{IdealRelevance}(q, r_i)} - 1}{\log_2(i + 1)}}$ & on hierarchical relevance. \\  \textbf{nDCG@k} &&                           \\ \hline
&& Balance between \\ \textbf{Hierarchical}         & $\frac{2 \cdot \text{Precision} \cdot \text{Recall}}{\text{Precision} + \text{Recall}}$& hierarchical precision \\  \textbf{F1 Score}&& and recall.                          \\ \hline
                                                       && Measures retrieval\\  \textbf{Hierarchical}         & $1 - \frac{\sum_{i=1}^k \text{Relevance}(q, r_i)}{k}$      & of irrelevant documents \\ \textbf{Severity} && in hierarchical setting.         \\ \hline
                                                
\textbf{Hierarchical} && Fraction of \\ 
                                             \textbf{False Positive}& $\frac{\text{Irrelevant hierarchical documents in top } k}{k}$ & irrelevant hierarchical \\ 
                                              \textbf{Rate@k}&& documents among top $k$. \\ \hline

\end{tabular}
\caption{Hierarchical Metrics for classification, retrieval and hallucination}
\label{tab:hierarchical_metrics}
\end{table}
%\paragraph{Hierarchical Metrics}

\begin{figure}[!ht]
    \centering
    % First subfigure
    \begin{subfigure}[b]{0.24\linewidth}
        \centering
        \includegraphics[width=\linewidth]{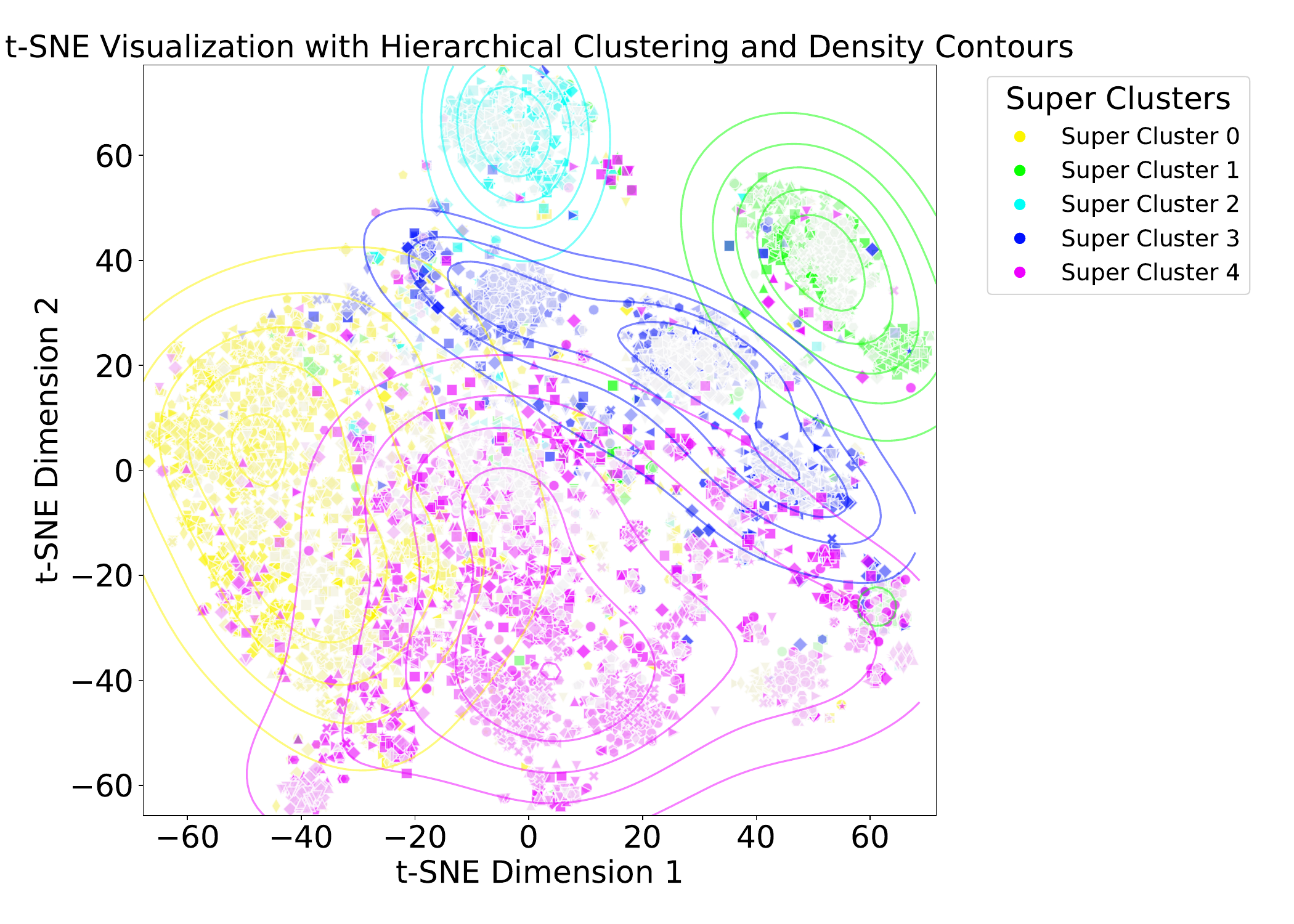} % Replace with your image
        \caption{}
        \label{fig:subfig1}
    \end{subfigure}
    % Second subfigure
    \begin{subfigure}[b]{0.24\linewidth}
        \centering
        \includegraphics[width=\linewidth]{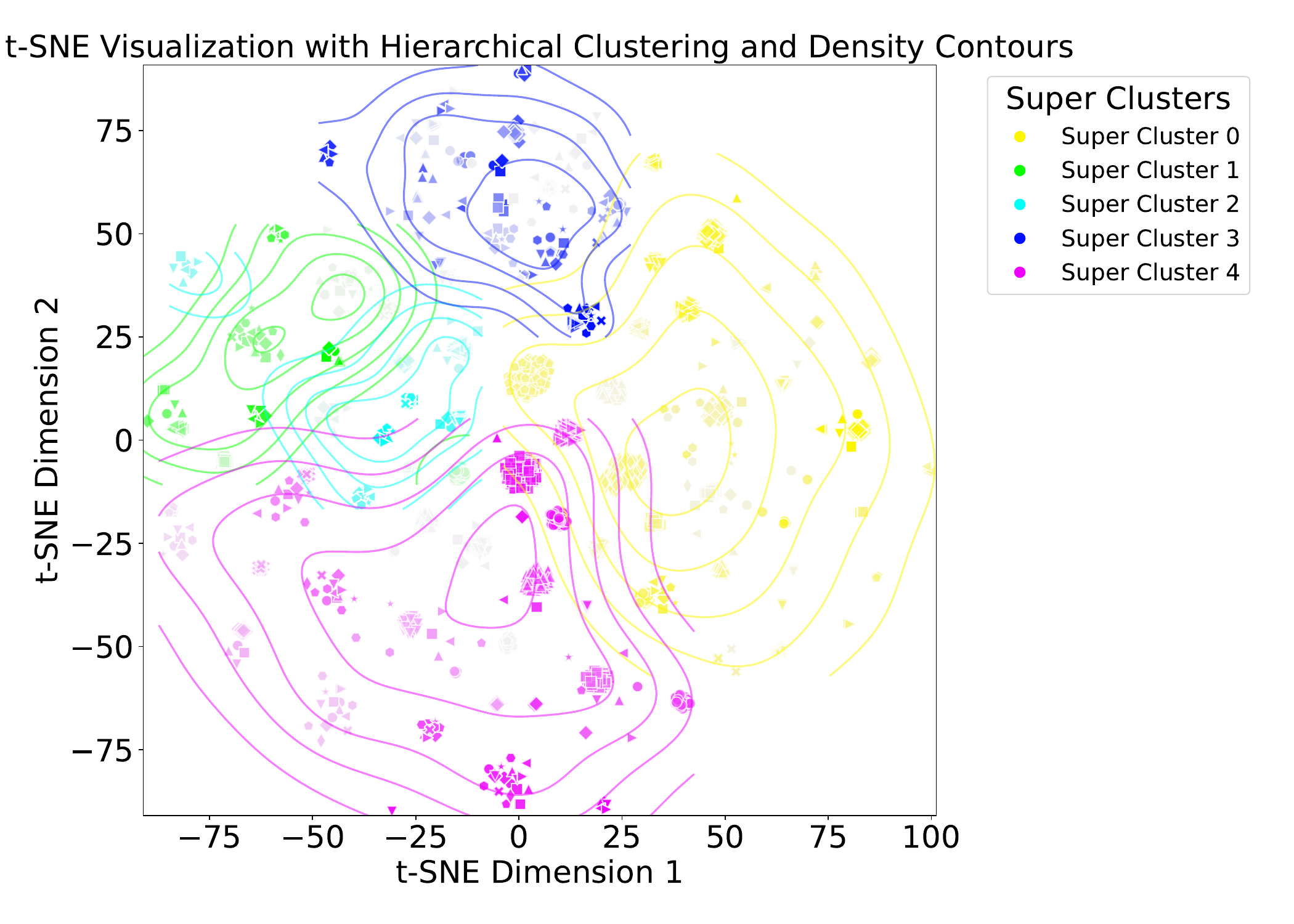} % Replace with your image
        \caption{}
        \label{fig:subfig2}
    \end{subfigure}
    % Third subfigure
    \begin{subfigure}[b]{0.24\linewidth}
        \centering
        \includegraphics[width=\linewidth]{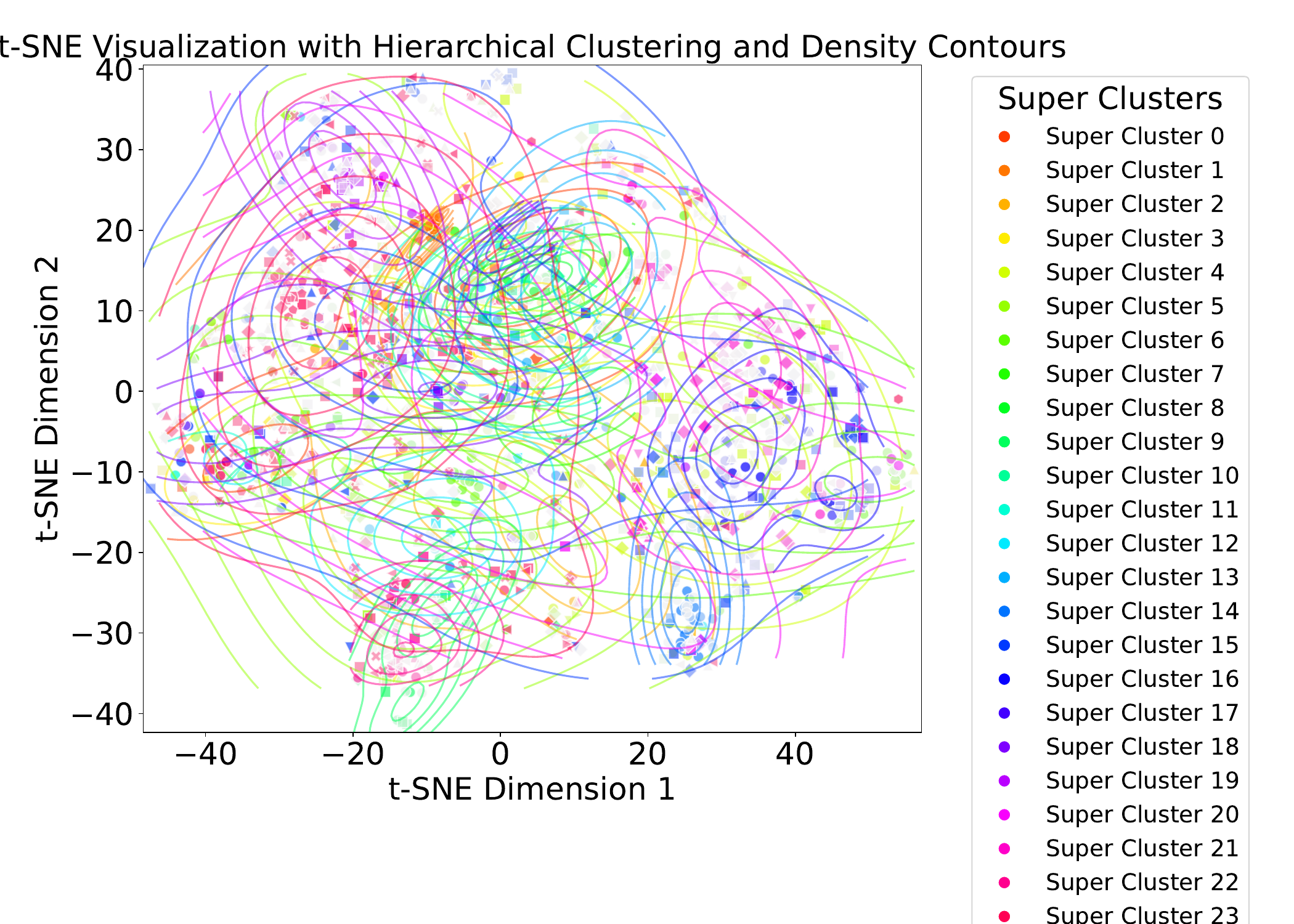} % Replace with your image
        \caption{}
        \label{fig:subfig3}
    \end{subfigure}
    % Fourth subfigure
    \begin{subfigure}[b]{0.24\linewidth}
        \centering
        \includegraphics[width=\linewidth]{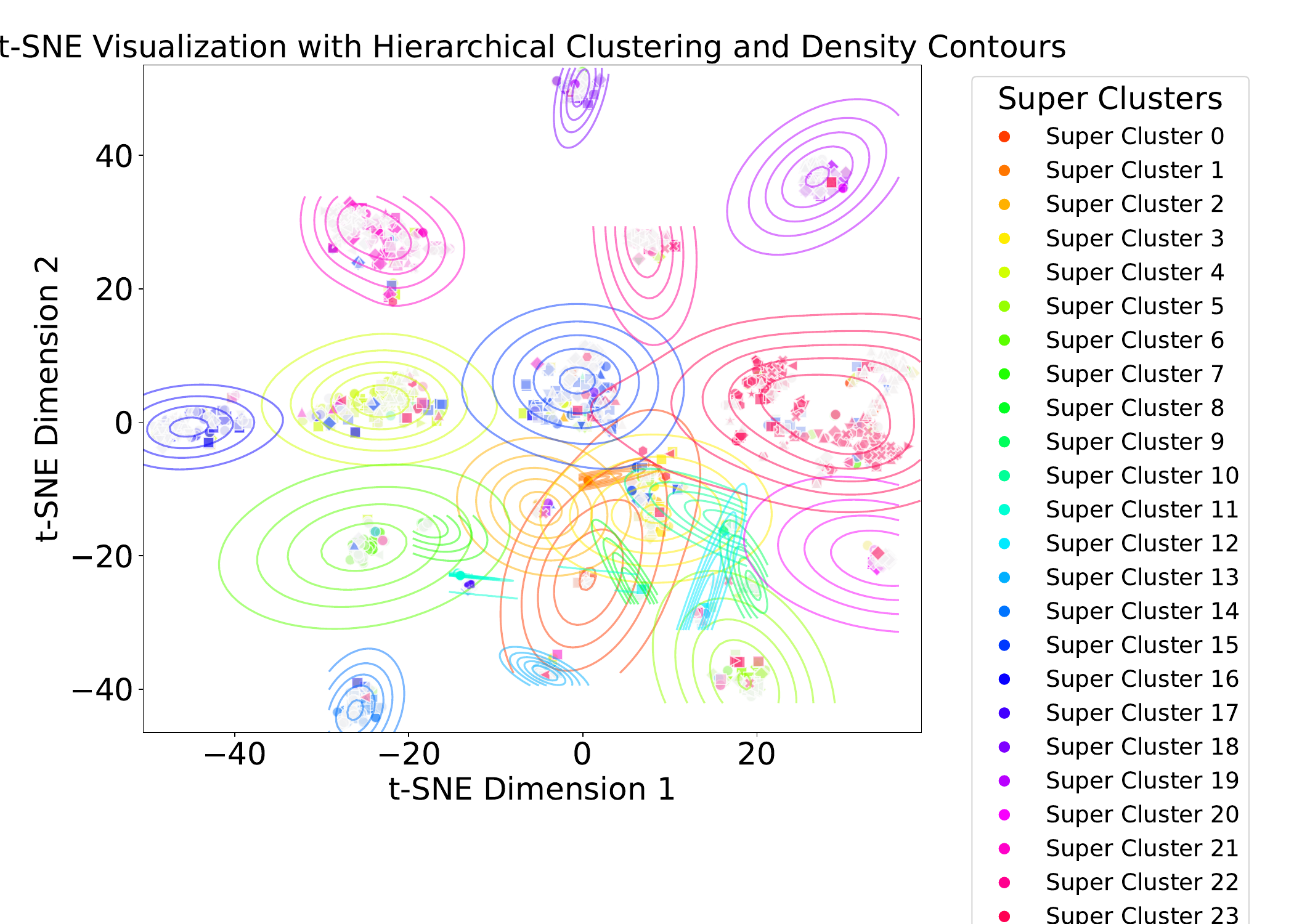} % Replace with your image
        \caption{}
       
    \end{subfigure}
\caption{Embedding visualizations for different datasets, projected using t-SNE for dimensionality reduction. The density contours represent the kernel density estimation (KDE) of the embeddings in the 2D space, highlighting the clustering structure. Subplots show the Material dataset (a) before and (b) after model alignment, and the Healthcare dataset (c) before and (d) after model alignment. The contours illustrate the density distribution of embeddings, showcasing the effect of alignment on cluster compactness and separation.}

  \label{fig:tsne}
\end{figure}

%\todo[]{for figure \ref{fig:side-by-side}, add few more senteces describing what we are looking at. how the embeddings are seperating the clusters better now}

\begin{table}[h!]
\centering
\caption{Performance Metrics Across Datasets (Healthcare, Materials, Cyber, Applied Mathematics) for Aligned and Non-aligned Embeddings for $k=10$}
\label{tab:results_all_datasets}
\resizebox{\textwidth}{!}{%
\begin{tabular}{|l|l|c|c|c|c|c|c|c|c|}
\hline
\multirow{2}{*}{\textbf{Task}} & \multirow{2}{*}{\textbf{Metric}} & \multicolumn{2}{c|}{\textbf{Healthcare}} & \multicolumn{2}{c|}{\textbf{Materials}} & \multicolumn{2}{c|}{\textbf{Cyber}} & \multicolumn{2}{c|}{\textbf{Applied Mathematics}} \\ \cline{3-10} 
                               &                                  & \textbf{Non-aligned} & \textbf{Aligned} & \textbf{Non-aligned} & \textbf{Aligned} & \textbf{Non-aligned} & \textbf{Aligned} & \textbf{Non-aligned} & \textbf{Aligned} \\ \hline
\multirow{3}{*}{Classification} 
& F1 Score                       & 0.5164               & 0.6588           & 0.6469               & 0.990           & 0.7130               & 0.8151           & 0.7541               & 0.8048           \\ 
& Precision                      & 0.5134               & 0.6590           & 0.6453               & 0.990           & 0.6975               & 0.8121           & 0.7415               & 0.8112           \\ 
& Recall                         & 0.5194               & 0.6586           & 0.6485               & 0.990           & 0.7293               & 0.8180           & 0.7672               & 0.7985           \\ \hline
\multirow{4}{*}{Retrieval}      
& Precision@k                    & 0.3103               & 0.4983           & 0.4787               & 0.9707           & 0.6397               & 0.7518           & 0.6576               & 0.7636           \\ 
& Recall@k                       & 0.0164               & 0.0290           & 0.0058               & 0.0116           & 0.0112               & 0.0133           & 0.0182               & 0.0212           \\ 
& MRR                            & 1.6259               & 2.2525           & 1.6541               & 2.9972           & 2.7538               & 3.1482           & 2.9065               & 3.2245           \\ 
& nDCG@k                         & 0.3752               & 0.5908           & 0.4982               & 0.990           & 0.6781               & 0.7908           & 0.7187               & 0.8280           \\ \hline
\multirow{2}{*}{Hallucination}  
& FPR@k                          & 0.9386               & 0.8771           & 0.8534               & 0.0878           & 0.7968               & 0.6236           & 0.8191               & 0.6529           \\ 
& Severity                       & 0.7306               & 0.5533           & 0.6041               & 0.0644           & 0.4402               & 0.3654           & 0.4119               & 0.3353           \\ \hline
\end{tabular}%
}
\end{table}

\subsection{Results}

Table~\ref{tab:results_all_datasets} summarizes the performance metrics for three datasets (Healthcare, Materials, Applied Mathematics, and Cybersecurity) across three tasks: classification, retrieval, and hallucination evaluation. The aligned model corresponds to the embedding model trained using the HEAL loss, whereas the non-aligned model corresponds to the original embedding model without HEAL-based training.  The metrics are reported for both non-aligned and aligned SciNCL embeddings, demonstrating the significant impact of HEAL on improving performance. 
Figure~\ref{fig:tsne} illustrates hierarchical embedding alignment achieved through HEAL training, resulting in well-separated super and sub-clusters for the Materials and Healthcare datasets which enhances the performance of downstream tasks.

First, we evaluate the performance on document-level tasks using hierarchical labels. Specifically, we assess the ability of the hierarchical classifier to predict hierarchical labels in the classification task. Additionally, we quantify the retrieval of documents from the same hierarchical category based on a query document to characterize retrieval accuracy and evaluate hallucinations.
The results presented in table~\ref{tab:results_all_datasets} demonstrate that HEAL significantly improves hierarchical classification metrics across all datasets. 
For the Healthcare dataset, the Hierarchical F1 Score improves from 0.5164 to 0.6588, reflecting a more accurate representation of hierarchical labels. Similarly, the Materials dataset achieves near perfect classification metrics (F1 Score, Precision, Recall = 0.99) with aligned embeddings, while the most challenging Healthcare dataset (4 depth cluster label) sees improvements in F1 Score from 0.5164 to 0.6588. In retrieval tasks, HEAL aligned embeddings consistently outperform non-aligned embeddings across all metrics. For the Healthcare dataset, Hierarchical MRR improves from 1.6259 to 2.2525, and nDCG@k increases from 0.3752 to 0.5908 where $k=10$, indicating better ranking and retrieval relevance. The Materials dataset achieves a dramatic increase in retrieval precision, with Precision@k rising from 0.4787 to 0.9707, while nDCG@k reaches 0.99, showcasing near-perfect retrieval performance. For the Cyber dataset, aligned embeddings yield an MRR improvement from 2.7538 to 3.1482 and a corresponding nDCG@k increase from 0.6781 to 0.7908. %These results highlight HEAL’s ability to produce embeddings that% better
%reflect hierarchical relationships, leading to accurate and relevant retrievals.
Hallucination metrics further underscore the superiority of HEAL. Aligned embeddings reduce hallucination rates significantly across all datasets. For the Healthcare dataset, FPR@k drops from 0.9386 to 0.8771, and severity decreases from 0.7306 to 0.5533, indicating fewer irrelevant or misleading retrievals. The Materials dataset shows the most striking improvement, with FPR@k reduced from 0.8534 to 0.0878 and severity declining from 0.6041 to 0.0644, nearly eliminating hallucination tendencies. For the Cyber dataset, aligned embeddings lower FPR@k from 0.7968 to 0.6236 and severity from 0.4402 to 0.3654.% These reductions in hallucination metrics illustrate the ability of HEAL to generate embeddings that minimize retrieval of irrelevant or misleading documents.%, ensuring higher-quality results.

Next, we evaluate the performance of aligned RAG in retrieving the correct documents for generated queries to augment the LLM and minimize hallucinations. From each test dataset, we randomly sampled 100 documents and generated 10 Q\&A pairs per document using the LLAMA-3.1 70B model, resulting in a total of 1,000 Q\&A pairs for each dataset. Each Q\&A pair was tagged with the corresponding document from which it was generated.
The prompt used for Q\&A generation was as follows: \textit{“First, provide a concise summary of the following abstract that emphasizes its key concepts and hierarchical relationships. Then, based on this summary, generate 10 unique, nuanced Q\&A pairs. Focus on creating questions that delve into specialized details of the hierarchical concepts discussed.”}
The generated queries were used to fetch documents via both aligned and unaligned models. We assessed the ability of each model to correctly retrieve the original document and evaluated the rank/order of retrieval. On average, the unaligned model achieved an MRR of 0.273 and a Recall@10 of 0.415. These metrics represent regular retrieval scores, not hierarchical scores. In contrast, the aligned model significantly improved performance, achieving an MRR of 0.514 and a Recall@10 of 0.731, demonstrating its superior ability to retrieve the correct set of documents.
Furthermore, when integrating RAG with LLAMA-3.1 70B for generating answers from the queries and retrieved documents, the unaligned model produced a ROUGE score of 0.42, while the aligned model achieved a ROUGE score of 0.68. This highlights the impact of alignment on improving the quality and relevance of generated responses.

\section{Conclusion}

In this work, we introduced HEAL, a novel framework for aligning embeddings in RAG systems through hierarchical fuzzy clustering and matrix factorization, integrated within a contrastive learning paradigm. HEAL effectively computes level-specific contrastive losses and applies hierarchical penalties to align embeddings with domain-specific structures, enhancing both retrieval relevance and classification performance. Experimental results across diverse domains — Healthcare, Materials Science, Cybersecurity, and Applied Mathematics — demonstrate HEAL's capability to significantly improve retrieval accuracy and mitigate hallucinations in LLM-based systems. By bridging hierarchical semantics with contrastive alignment, HEAL establishes itself as a versatile and robust tool for advancing RAG methodologies, enabling more precise, reliable, and domain-adaptive applications of large language models.

% References

\bibliographystyle{plainnat}
\bibliography{references}

\end{document}